\documentclass[aps,showpacs,prd,twocolumn]{revtex4-1}
\usepackage{amsfonts}
\usepackage{amssymb}
\usepackage{amsmath}
\usepackage{color}
\usepackage[usenames,dvipsnames]{xcolor}
\usepackage{float}
\usepackage{accents}
\usepackage{graphicx}
\usepackage{epstopdf}
\usepackage[colorlinks=true,pdfstartview=FitV,linkcolor=blue,citecolor=blue,urlcolor=blue,breaklinks=true]{hyperref}

\begin{document}

\title{Topological first-order solitons in a gauged $CP(2)$ model with the
Maxwell-Chern-Simons action}
\author{R. Casana}
\email{rodolfo.casana@gmail.com }
\author{N. H. Gonzalez-Gutierrez}
\email{neyver.hgg@gmail.com }
\author{E. da Hora}
\email{edahora.ufma@gmail.com}
\affiliation{$\color{blue}{}^{*\,\dag}$Departamento de F\'{\i}sica, Universidade Federal
do Maranh\~{a}o, 65080-805, S\~{a}o Lu\'{\i}s, Maranh\~{a}o, Brazil,}
\affiliation{$\color{blue}{}^{\ddag}$Coordenadoria Interdisciplinar de Ci\^{e}ncia e
Tecnologia, Universidade Federal do Maranh\~{a}o, {65080-805}, S\~{a}o Lu%
\'{\i}s, Maranh\~{a}o, Brazil.}

\begin{abstract}
We verify the existence of radially symmetric first-order solitons in a
gauged $CP(2)$ scenario in which the dynamics of the Abelian gauge field is
controlled by the Maxwell-Chern-Simons action. We implement the standard
Bogomol'nyi-Prasad-Sommerfield (BPS) formalism, from which we obtain a
well-defined lower bound for the corresponding energy (i.e. the Bogomol'nyi
bound) and the first-order equations saturating it. We solve these
first-order equations numerically by means of the finite-difference scheme,
therefore obtaining regular solutions of the effective model, their energy
being quantized according the winding number rotulating the final
configurations, as expected. We depict the numerical solutions, whilst
commenting on the main properties they engender.
\end{abstract}

\maketitle

\section{Introduction}

Time-independent solutions of highly nonlinear equations are of great
importance and interest in many areas of physics and mathematics \cite{n5}. {%
In field theory such a highly nonlinear equations arise naturally, i.e, the
second-order Euler-Lagrange equations, which can be quite hard to solve}.
However, under very special circumstances, genuine field solutions can be
obtained via a particular set of two coupled first-order differential
equations, namely the Bogomol'nyi-Prasad-Sommerfield (BPS) ones \cite{BPS},
the resulting configurations minimizing the energy of the overall system.

In the context of gauged (2+1)-dimensional models, these first-order
solutions are called vortices. In particular, magnetic vortices were
verified to occur in the Maxwell-Higgs electrodynamics \cite{n1}. Also, it
was demonstrated that electrically charged vortices emerge from both the
Chern-Simons-Higgs \cite{cshv} and the Maxwell-Chern-Simons-Higgs scenarios 
\cite{mcshv}.

Moreover, other important examples of first-order solutions include the ones
arising from nonlinear sigma models (NL$\sigma $M) \cite{polyakov} in the
presence of a gauge field, which have been widely applied in the study of
different aspects of field theory and condensed matter physics \cite{CMP}.

In this sense, the existence of topological solitons in a $O(3)$ nonlinear
sigma model endowed by the Maxwell action was demonstrated in \cite%
{schroers,mukherjee2}. Moreover, in the Refs. \cite{ghosh,mukherjee1}, the
authors studied the $O(3)$ nonlinear sigma model gauged by the Chern-Simons
term, establishing the existence of both topological and nontopological
configurations. The gauged $O(3)$ sigma model with the gauge field dynamics
ruled by both the Maxwell and the Chern-Simons terms was also studied \cite%
{sigmaMCSH1,sigmaMCSH2}.

Topological solitons also appear in the $CP(N-1)$ models whose importance is
due to the fact that they present some fundamental properties (such as
asymptotic freedom, confinement, nontrivial vacuum structure giving rise to
stable instantons, etc.) typically inherent to the Yang-Mills theories \cite%
{cpn-1.1,cpn-1.2,cpn-1.3,cpn-1.4}.

In a recent work, the existence of first-order vortices in a gauged $CP(2)$
model whose gauge field is ruled by the Maxwell action was proposed \cite%
{loginov}. This hypothesis was confirmed in the Ref. \cite{casana}, where
the first-order formalism was implemented in a clear way, giving rise to a
well-defined lower-bound for the total energy (i.e.the Bogomol'nyi bound)
and to the corresponding first-order differential equations.

In the sequel, some of us have considered a gauged $CP(2)$ scenario endowed
by the Maxwell term multiplied by a nontrivial dielectric function, the
resulting noncanonical model supporting nontopological first-order vortices
with no quantized magnetic flux \cite{lima}. Furthermore, some of us have
also calculated the topological first-order vortices inherent to a gauged $%
CP(2)$ model in the presence of the Chern-Simons term (instead of the
Maxwell one), see the Ref. \cite{caduCS}.

In this context, the aim of the present manuscript is to investigate the
existence of first-order solitons arising from a gauged $CP(2)$ model in
which the dynamics of the gauge sector is ruled simultaneously by both the
Maxwell and the Chern-Simons terms. Therefore, in order to introduce our
results, this work is organized as follows: in the next Sec. II, we define
the gauged Maxwell-Chern-Simons $CP(N-1)$ model, focusing our attention on
those time-independent solitons possessing radial symmetry. We then
particularize our investigation to the case $N=3$, from which we develop the
corresponding first-order framework via the usual prescription (i.e.
requiring the minimization the total energy), this way finding the resulting
lower-bound for the energy itself and the first-order equations saturating
it. In the Sec. III, we solve the first-order equations numerically by means
of the finite-difference algorithm, from which we depict the numerical
solutions, whilst commenting on the main properties they engender. We end
our work in the Section IV, in which we point out our final observations and
perspectives regarding future contributions.

\section{The overall model}

We begin our manuscript by considering the gauged $CP(N-1)$ model \cite%
{loginov} in the presence of the usual Chern-Simons term, the corresponding
Lagrange density standing for 
\begin{equation}
\mathcal{L}_{0}=\mathcal{L}_{MCS}+\left( P_{ab}D_{\mu }\phi _{b}\right)
^{\ast }P_{ac}D^{\mu }\phi _{c}-U\left( \left\vert \phi \right\vert \right) 
\text{,}  \label{ai1}
\end{equation}%
where%
\begin{equation}
\mathcal{L}_{MCS}=\mathcal{L}_{M}+\mathcal{L}_{CS}\text{,}
\end{equation}%
with%
\begin{equation}
\mathcal{L}_{M}=-\frac{1}{4}F_{\mu \nu }F^{\mu \nu }\text{ \ and \ }\mathcal{%
L}_{CS}=-\frac{\kappa }{4}\epsilon ^{\rho \mu \nu }A_{\rho }F_{\mu \nu }%
\text{.}
\end{equation}%
Here, $F_{\mu \nu }=\partial _{\mu }A_{\nu }-\partial _{\nu }A_{\mu}$ is the
standard electromagnetic field strength tensor and $D_{\mu}\phi_{a}=
\partial_{\mu} \phi _{a}-igA_{\mu }Q_{ab}\phi _{b}$ stands for the
corresponding covariant derivative. Also, $P_{ab}=\delta _{ab}-h^{-1} \phi
_{a}\phi _{b}^{\ast }$ is a projection operator introduced for the sake of
convenience, whilst $Q_{ab}$ represents a real and diagonal charge matrix.
In the present case, the $CP(N-1)$ field is assumed to satisfy $%
\phi_{a}^{\ast} \phi _{a}=h.$

The corresponding Euler-Lagrange equations for the gauge and scalar sectors
are, respectively,%
\begin{equation}
\partial _{\mu }F^{\mu \rho }-\frac{\kappa }{2}\epsilon ^{\rho \mu \nu
}F_{\mu \nu }=J^{\rho }
\end{equation}%
and%
\begin{equation}
2P_{ad}D_{\mu }\left( P_{dc}D^{\mu }\phi _{c}\right) -P_{ad}D_{\mu }D^{\mu
}\phi _{d}=-P_{ad}\frac{\partial U}{\partial \phi _{d}^{\ast }}\text{,}
\end{equation}
where 
\begin{equation}
J^{\mu }=ig\left[ \left( P_{ab}D^{\mu }\phi _{b}\right) ^{\ast }\left(
P_{ac}Q_{cd}\phi _{d}\right) -\text{h.c.}\right]
\end{equation}%
stands for the current 4-vector (conserved). Here, h.c. means Hermitian
conjugate.

We look for the time-independent solutions arising from (\ref{ai1}). In this
sense, it is instructive to write down the Gauss law for static
configurations, i.e., (here, $B=F_{12}$ is the magnetic field) 
\begin{equation}
\partial _{j}\partial _{j}A_{0}+\kappa B=2g^{2}A^{0}\left( P_{ac}Q_{cd}\phi
_{d}\right) ^{\ast }P_{ab}Q_{bm}\phi _{m}\text{,}  \label{gauss1}
\end{equation}%
the corresponding solutions possessing both magnetic and electric fields.

In addition, the time-independent Amp\`{e}re's law can be written as%
\begin{equation}
\partial _{k}B+\kappa \partial _{k}A_{0}=-\epsilon _{kj}J_{j}\text{,}
\end{equation}%
with%
\begin{equation}
J_{k}=ig\left[ \left( P_{ab}D_{k}\phi _{b}\right) ^{\ast }\left(
P_{ac}Q_{cd}\phi _{d}\right) -\text{h.c.}\right] \text{,}
\end{equation}%
the equation of motion for the static scalar sector reading 
\begin{eqnarray}
&&2P_{ad}D_{k}\left( P_{dc}D_{k}\phi _{c}\right) -P_{ad}D_{k}D_{k}\phi _{d} 
\notag \\[0.15cm]
&&-P_{ad}\frac{\partial U}{\partial \phi _{d}^{\ast }}+2g^{2}\left(
A_{0}\right) ^{2}P_{ad}Q_{db}P_{bc}Q_{ce}\phi _{e}  \notag \\[0.15cm]
&&-g^{2}\left( A_{0}\right) ^{2}P_{ad}Q_{db}Q_{be}\phi _{e}=0\text{.}
\end{eqnarray}

It can be shown that the Lagrange density (\ref{ai1}) does not support
solitonic configurations satisfying a well-defined first-order framework.
However, the existence of planar first-order solitons carrying both magnetic
and electric fields becomes possible via the introduction of a neutral
scalar field $\Psi $ into the Lagrange density (\ref{ai1}), the resulting
model being described by 
\begin{eqnarray}
\mathcal{L} &=&\mathcal{L}_{MCS}+\left( P_{ab}D_{\mu }\phi _{b}\right)
^{\ast }P_{ac}D^{\mu }\phi _{c}+\frac{1}{2}\partial _{\mu }\Psi \partial
^{\mu }\Psi  \notag \\[0.08in]
&&-g^{2}\Psi ^{2}\left( P_{ab}Q_{bd}\phi _{d}\right) ^{\ast
}P_{ac}Q_{ce}\phi _{e}-U\left( |\phi |,\Psi \right) \text{,}  \label{ai2a}
\end{eqnarray}%
$U\left( |\phi |,\Psi \right) $ standing for the potential describing the
scalar-matter self-interaction (to be determined later).

Here, it is worthwhile to point out that the introduction of such a neutral
field is a well-established prescription supporting the consistent study of
first-order configurations in the presence of the Chern-Simons term. This
prescription was firstly used in the context of the
Maxwell-Chern-Simons-Higgs theories \cite{lee,bolog}, the main motivation
coming from supersymmetric arguments. It was also implemented successfully
in connection to nonlinear sigma models \cite{sigmaMCSH1,sigmaMCSH2}, being
also used to describe charged solitons arising from Lorentz-violating
scenarios \cite{casana1,Guillermo,Claudio,Claudio1}.

Along of the remain of this manuscript, we consider the $N=3$ case, this way
reducing our study to the $CP(2)$ scenario. We then focus our attention on
those time-independent solutions presenting radial symmetry which are
implemented via the standard \textit{Ansatz}%
\begin{equation}
A_{0}=A_{0}\left( r\right) \text{ \ and \ }A_{i}=-\frac{1}{gr^{2}}\epsilon
_{ij}x_{j}A\left( r\right) \text{,}  \label{li0}
\end{equation}%
\begin{equation}
\left( 
\begin{array}{c}
\phi _{1} \\ 
\phi _{2} \\ 
\phi _{3}%
\end{array}%
\right) =h^{1/2}\left( 
\begin{array}{c}
e^{in_{1}\theta }\sin \alpha \left( r\right) \cos \beta \left( r\right) \\ 
e^{in_{2}\theta }\sin \alpha \left( r\right) \sin \beta \left( r\right) \\ 
e^{in_{3}\theta }\cos \alpha \left( r\right)%
\end{array}%
\right) \text{,}  \label{li1}
\end{equation}%
where $\epsilon _{ij}$ stands for the bidimensional Levi-Civita tensor (with 
$\epsilon _{12}=+1$), $x_{i}=r(\cos \theta ,\sin \theta )$ is the position
vector (written in polar coordinates) and the integers $n_{1}$, $n_{2}$ and $%
n_{3}$ represent the winding numbers (vorticities) of the corresponding
scalar solutions.

\begin{figure}[tbp]
\centering\includegraphics[width=8.5cm]{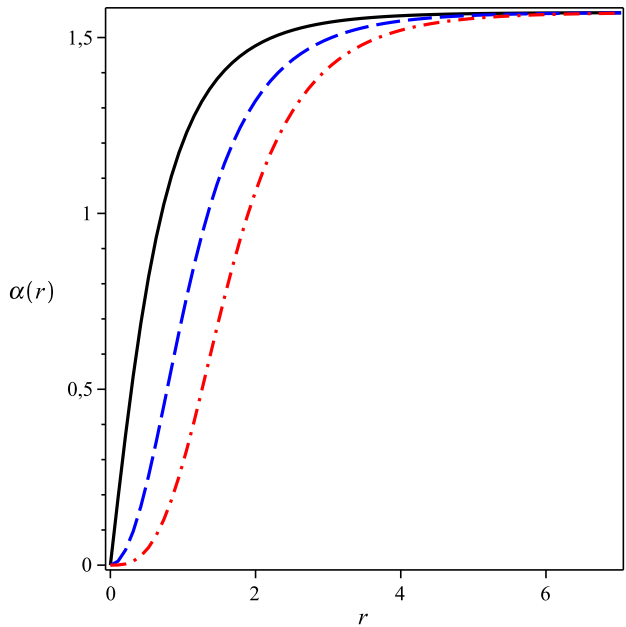}
\caption{Numerical solutions to the profile function $\protect\alpha (r)$
for $n=1$ (solid black line), $n=2$ (dashed blue line) and $n=3$ (dotted red
line). Here, $h=\protect\kappa =1$ and $g=2$.}
\end{figure}

Now, it is important to highlight that, concerning the combination between
the charge matrix $Q_{ab}$ and the winding numbers $n_{1}$, $n_{2}$ and $%
n_{3}$, there are two possible choices supporting the existence of
topological solitons: (i) $Q=\lambda _{3}/2$ and $n_{1}=-n_{2}=n$, and (ii) $%
Q=\lambda _{8}/2$ and $n_{1}=n_{2}=n$ (both ones with $n_{3}=0$, $\lambda
_{3}$ and $\lambda _{8}$ being the diagonal Gell-Mann matrices, i.e., $%
\lambda _{3}=\mbox{diag}(1,-1,0)$ and $\sqrt{3}\lambda _{8}=\mbox{diag}%
(1,1,-2)$). Nevertheless, in \cite{loginov}, the author have demonstrated
that these two combinations simply mimic each other, this way existing only
one effective scenario. Therefore, in this manuscript, we investigate only
the case defined by $n_{1}=-n_{2}=n$, $n_{3}=0$ and 
\begin{equation}
Q_{ab}=\frac{1}{2}\lambda _{3}=\frac{1}{2}\text{diag}\left( 1,-1,0\right) 
\text{.}
\end{equation}

Furthermore, given the Ansatz (\ref{li0}) and (\ref{li1}) and the
conventions stated above, one gets that nonsingular solutions possessing
finite-energy are attained by those profile functions $\alpha (r)$, $A(r)$
and $A_{0}(r)$\ obeying the boundary conditions%
\begin{equation}
\alpha (r=0)=0\text{, \ }A(r=0)=0\text{, \ }A_{0}^{\prime }(r=0)=0\text{,}
\label{ai00}
\end{equation}%
and%
\begin{equation}
\alpha (r\rightarrow \infty )\rightarrow \frac{\pi }{2}\text{, \ }%
A(r\rightarrow \infty )\rightarrow 2n\text{, \ }A_{0}(r\rightarrow \infty
)\rightarrow 0\text{,}  \label{ai01}
\end{equation}%
where prime denotes the derivative with respect to the radial coordinate $r$%
. Here, it is important to point out that the boundary conditions for $%
\alpha (r)$ and $A(r)$ are well-established in the literature \cite{loginov}%
. In addition, we verify the conditions for the electric potential $A_{0}(r)$
in the Sec. III below.

\begin{figure}[tbp]
\centering\includegraphics[width=8.5cm]{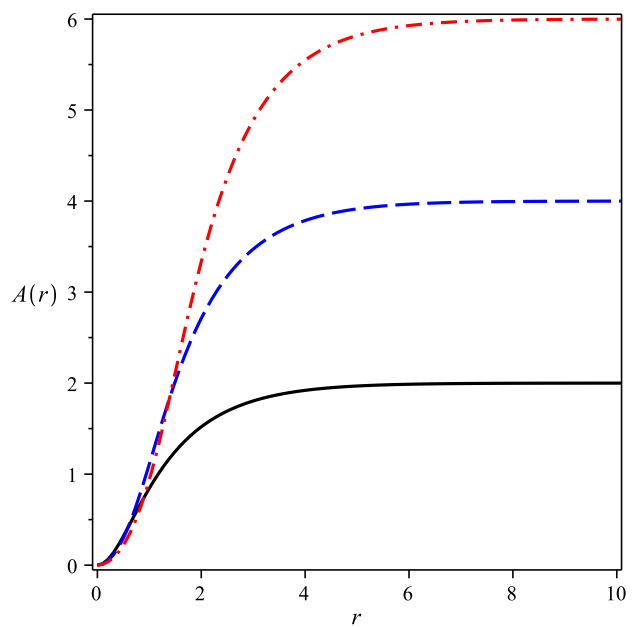}
\caption{Numerical solutions to the profile function $A(r)$. Conventions as
in the Fig. 1. The solutions reach the asymptotic value $A(r\rightarrow
\infty )\rightarrow 2n$ in a monotonic way, as expected.}
\end{figure}

It is also instructive to write down the equation of motion for the
additional profile function $\beta \left( r\right) $, i.e. (here, we have
defined $H_{\pm }(r)=\left( A_{0}\right) ^{2}\pm \Psi ^{2}$, for the sake of
simplicity)%
\begin{eqnarray}
&&\frac{d^{2}\beta }{dr^{2}}+\left( \frac{1}{r}+2\cot \alpha \frac{d\alpha }{%
dr}\right) \frac{d\beta }{dr}+\frac{g^{2}}{4}H_{-}(r)\sin ^{4}\alpha \sin
\left( 4\beta \right)  \notag \\[0.06in]
&&-\frac{\sin ^{2}\alpha \sin \left( 4\beta \right) }{r^{2}}\left( n-\frac{A%
}{2}\right) ^{2}=0\text{,}  \label{beta0}
\end{eqnarray}%
whose two possible solutions are%
\begin{equation}
\beta _{1}=\frac{\pi }{4}+\frac{\pi }{2}k\text{ \ or \ }\beta _{2}=\frac{\pi 
}{2}k\text{,}  \label{betha}
\end{equation}%
with $k\in \mathbb{Z}$. These solutions define \emph{a priori} two different
scenarios. However, the expressions arising from $\beta (r)=\beta _{2}$ can
be obtained directly from those inherent to $\beta (r)=\beta _{1}$ via the
transformations $\alpha \rightarrow \alpha /2$ and $h\rightarrow 4h$. Hence,
we conclude that these two a priori different solutions are
phenomenologically equivalent, this way existing only one effective
scenario. So, in what follows, we only consider the case $\beta (r)=\beta
_{1}$.

\begin{figure}[tbp]
\centering\includegraphics[width=8.5cm]{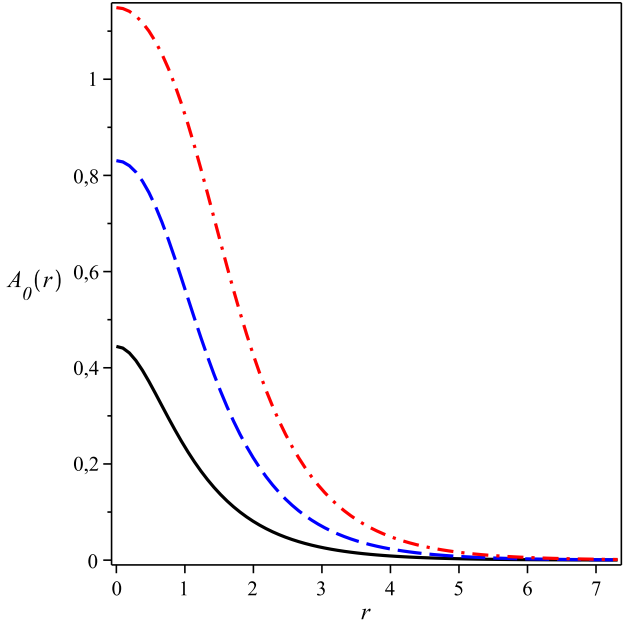}
\caption{Numerical solutions to the gauge function $A_{0}(r)$. Conventions
as in the Fig. 1, $A_{0}(0)\equiv A_{0}(r=0)$ increasing with the vorticity.}
\end{figure}

We now look for first-order differential equations supporting radially
symmetric solutions of the model (\ref{ai2a}). As it is usual, we proceed
the minimization of the corresponding energy, the starting-point being the
energy-momentum tensor inherent to (\ref{ai2a}), i.e., 
\begin{eqnarray}
\mathcal{T}_{\rho \sigma } &=&-F_{\rho \mu }F_{\sigma }{}^{\mu }+\partial
_{\rho }\Psi \partial _{\sigma }\Psi +\left( P_{ab}D_{\rho }\phi _{b}\right)
^{\ast }P_{ac}D_{\sigma }\phi _{c}  \notag \\[0.08in]
&&+\left( P_{ab}D_{\sigma }\phi _{b}\right) ^{\ast }P_{ac}D_{\rho }\phi
_{c}-g_{\rho \sigma }\mathcal{L}_{N}\text{,}  \label{ai3a}
\end{eqnarray}%
where 
\begin{eqnarray}
\mathcal{L}_{N} &=&\mathcal{L}_{M}+\left( P_{ab}D_{\mu }\phi _{b}\right)
^{\ast }P_{ac}D^{\mu }\phi _{c}+\frac{1}{2}\partial _{\mu }\Psi \partial
^{\mu }\Psi  \notag \\[0.08in]
&&-g^{2}\Psi ^{2}\left( P_{ab}Q_{bd}\phi _{d}\right) ^{\ast
}P_{ac}Q_{ce}\phi _{e}-U
\end{eqnarray}%
stands for the nontopological part of the Lagrange density (\ref{ai2a}).
Thus, one gets the radially symmetric energy density to be 
\begin{eqnarray}
\varepsilon &=&\frac{1}{2}B^{2}+U+\frac{1}{2}\left( \frac{dA_{0}}{dr}\right)
^{2}+\frac{1}{2}\left( \frac{d\Psi }{dr}\right) ^{2}  \notag \\[0.06in]
&&+h\left[ \left( \frac{d\alpha }{dr}\right) ^{2}+\frac{W}{4r^{2}}\left(
2n-A\right) ^{2}\sin ^{2}\alpha \right]  \notag \\[0.06in]
&&+h\left[ \left( \frac{d\beta }{dr}\right) ^{2}+\frac{g^{2}W}{4}H_{+}(r)%
\right] \sin ^{2}\alpha \text{,}  \label{ee0}
\end{eqnarray}%
where we have introduced the auxiliary function 
\begin{equation}
W\left( \alpha ,\beta \right) =1-\sin ^{2}\alpha \cos ^{2}\left( 2\beta
\right) \text{,}
\end{equation}%
the magnetic field $B(r)$\ being given by%
\begin{equation}
B(r)=\frac{1}{gr}\frac{dA}{dr}\text{,}
\end{equation}%
according the \emph{Ansatz} (\ref{li0}) and (\ref{li1}). Here, $U=U(\alpha
,\Psi )$ represents self-interacting potential.

\begin{figure}[tbp]
\centering\includegraphics[width=8.5cm]{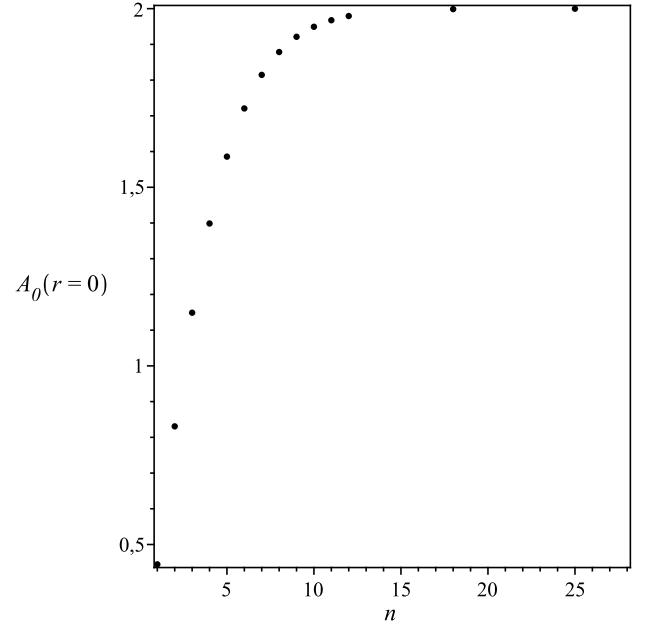}
\caption{The dependence of $A_{0}(0)\equiv A_{0}(r=0)$ with the vorticity:
given $h=\protect\kappa =1$ and $g=2$, the scalar potential reaches $A_{0}^{%
\text{max}}(r=0)=2$ for sufficiently large values of $n$.}
\end{figure}

We now choose the particular solution 
\begin{equation}
\beta \left( r\right) =\beta _{1}=\frac{\pi }{4}+\frac{\pi }{2}k\text{,}
\end{equation}%
from which we get $W(\alpha ,\beta =\beta _{1})=1$, the energy density (\ref%
{ee0})being reduced to 
\begin{eqnarray}
\varepsilon &=&\frac{1}{2}B^{2}+U+h\left( \frac{d\alpha }{dr}\right) ^{2}+%
\frac{h}{4r^{2}}\left( 2n-A\right) ^{2}\sin ^{2}\alpha  \notag \\
&&+\frac{1}{2}\left( \frac{dA_{0}}{dr}\right) ^{2}+\frac{1}{2}\left( \frac{%
d\Psi }{dr}\right) ^{2}+\frac{g^{2}h}{4}H_{+}\sin ^{2}\alpha \text{,}
\label{ee1}
\end{eqnarray}%
which can be rewritten as 
\begin{eqnarray}
\varepsilon &=&\frac{1}{2}\left( B\mp \sqrt{2U}\right) ^{2}+h\left[ \frac{%
d\alpha }{dr}\mp \frac{\left( 2n-A\right) }{2r}\sin \alpha \right] ^{2} 
\notag \\[0.15cm]
&&+\frac{1}{2}\left( \frac{d\Psi }{dr}\pm \frac{dA_{0}}{dr}\right) ^{2}+%
\frac{g^{2}h}{4}\left( \Psi \pm A_{0}\right) ^{2}\sin ^{2}\alpha  \notag \\
&&\pm B\sqrt{2U}\pm h\frac{\left( 2n-A\right) }{r}\frac{d\alpha }{dr}\sin
\alpha  \notag \\[0.15cm]
&&\mp \frac{d\Psi }{dr}\frac{dA_{0}}{dr}\mp \frac{g^{2}h}{2}\Psi A_{0}\sin
^{2}\alpha \text{.}  \label{ee2}
\end{eqnarray}

\begin{figure}[tbp]
\centering\includegraphics[width=8.5cm]{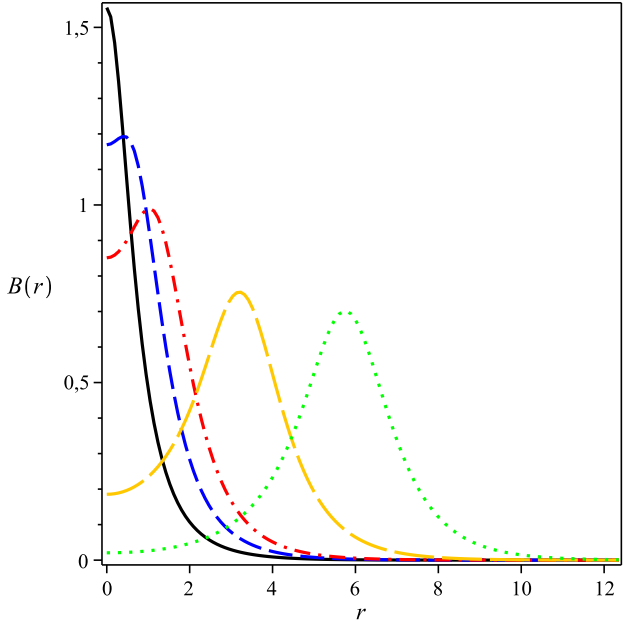}
\caption{Numerical solutions to the magnetic field $B(r)$. Conventions as in
the Fig. 1, the long-dashed orange line (dotted green line) representing the
solution for $n=7$ ($n=12$).}
\end{figure}

Here, it is useful to note that, in view of (\ref{li0}) and (\ref{li1}), the
time-independent Gauss law (\ref{gauss1}) can be written in the form (note
that such equation remains the same after the introduction of the neutral
scalar field $\Psi $), 
\begin{equation}
\frac{d^{2}A_{0}}{dr^{2}}+\frac{1}{r}\frac{dA_{0}}{dr}+\kappa B=\frac{g^{2}h%
}{2}A_{0}\sin ^{2}\alpha \text{.}  \label{gg1}
\end{equation}%
It allows the last term in (\ref{ee2}) can be rewritten as 
\begin{equation}
\frac{g^{2}h}{2}A_{0}\Psi \sin ^{2}\alpha =\Psi \frac{d^{2}A_{0}}{dr^{2}}+%
\frac{\Psi }{r}\frac{dA_{0}}{dr}+\kappa \Psi B\text{,}
\end{equation}%
consequently, the energy density (\ref{ee2}) being reduced to be 
\begin{eqnarray}
\varepsilon &=&\frac{1}{2}\left( B\mp \sqrt{2U}\right) ^{2}+h\left[ \frac{%
d\alpha }{dr}\mp \frac{\left( 2n-A\right) }{2r}\sin \alpha \right] ^{2} 
\notag \\[0.15cm]
&&+\frac{1}{2}\left( \frac{d\Psi }{dr}\pm \frac{dA_{0}}{dr}\right) ^{2}+%
\frac{1}{4}hg^{2}\left( \Psi \pm A_{0}\right) ^{2}\sin ^{2}\alpha  \notag \\%
[0.15cm]
&&\pm B\left( \sqrt{2U}-hg\cos \alpha -\kappa \Psi \right)  \notag \\[0.15cm]
&&\mp \frac{h}{r}\frac{d}{dr}\left[ \left( 2n-A\right) \cos \alpha \right]
\mp \frac{1}{r}\frac{d}{dr}\left( r\Psi \frac{dA_{0}}{dr}\right) \text{.}
\label{ee3}
\end{eqnarray}%
Due to the boundary conditions, the last term in (\ref{ee3}) (i.e. the total
derivative $r^{-1}\left( r\Psi A_{0}^{\prime }\right) ^{\prime }$) gives
null contribution to the total energy, so we neglect it. However, the
penultimate total derivative provides a nonvanishing contribution to the
overall energy.

The potential $U(\vert \phi \vert ,\Psi) $ is determined by choosing 
\begin{equation}
U\left( \alpha ,\Psi \right) =\frac{1}{2}\left( gh\cos \alpha +\kappa \Psi
\right) ^{2}\text{,}  \label{pot1}
\end{equation}
\textbf{so}, the resulting total energy being given by 
\begin{eqnarray}
E &=&\pm 4\pi hn+2\pi \int_{0}^{\infty }\!\!r\left\{ \frac{1}{2}\left[ \!%
\frac{{}}{{}}B\mp \left( gh\cos \alpha +\kappa \Psi \right) \right]
^{2}\right.  \notag \\[0.06in]
&&\ \ \ \ \ \ +\frac{1}{2}\left( \frac{d\Psi }{dr}\pm \frac{dA_{0}}{dr}%
\right) ^{2}+\frac{g^{2}h}{4}\left( \Psi \pm A_{0}\right) ^{2}\sin ^{2}\alpha
\notag \\[0.06in]
&&\ \ \ \ \ \ +\left. h\left[ \frac{d\alpha }{dr}\mp \frac{\left(
2n-A\right) }{2r}\sin \alpha \right] ^{2}\right\} dr\text{.}  \label{ee4}
\end{eqnarray}

\begin{figure}[tbp]
\centering\includegraphics[width=8.5cm]{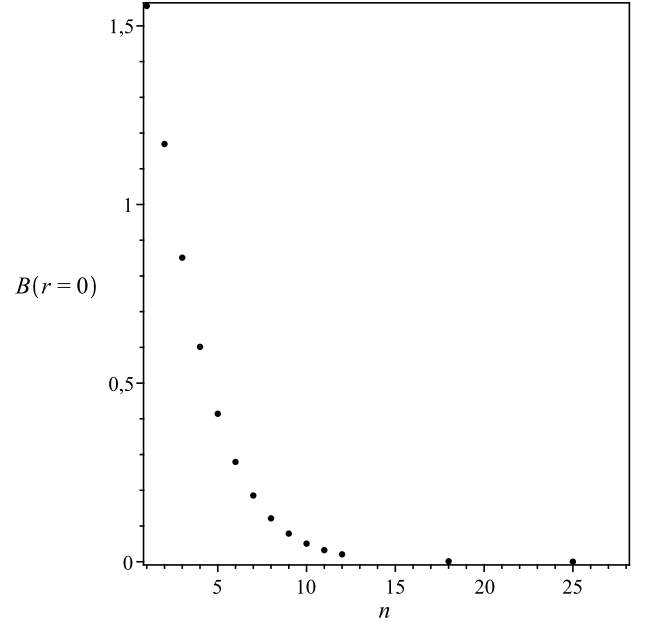}
\caption{The dependence of $B(0)$ with the vorticity: the difference $%
b_0=gh- \protect\kappa A_{0}(0)$ vanishes for sufficiently large values of $%
n $.}
\end{figure}

The expression in (\ref{ee4}) reveals that the energy of the effective model
supports a well-defined lower-bound given by 
\begin{equation}
E_{bps}=\pm 4\pi hn=\mp 2\pi \int_{0}^{\infty }dr~r\frac{h}{r}\frac{d}{dr}%
\left[ \left( 2n-A\right) \cos \alpha \right] \text{,}  \label{lwb-1}
\end{equation}%
which is attained by those fields satisfying the set of differential
equations 
\begin{equation}
\frac{d\alpha }{dr}=\pm \frac{\left( 2n-A\right) }{2r}\sin \alpha \text{,}
\end{equation}%
\begin{equation}
B=\pm \left( hg\cos \alpha +\kappa \Psi \right) \text{,}
\end{equation}%
\begin{equation}
\frac{d\Psi }{dr}=\mp \frac{dA_{0}}{dr}\text{ \ and \ }\Psi =\mp A_{0}\text{,%
}
\end{equation}%
which therefore stand for the first-order equations inherent to the
effective radially symmetric model.

Here, it is interesting to highlight that the last two equations are
automatically satisfied by $\Psi =\mp A_{0}$, the resulting configurations
being then obtained via the two remaining first-order equations, i.e. 
\begin{equation}
\frac{d\alpha }{dr}=\pm \frac{\left( 2n-A\right) }{2r}\sin \alpha \text{,}
\label{bps11}
\end{equation}%
\begin{equation}
B=\frac{1}{gr}\frac{dA}{dr}=\pm hg\cos \alpha -\kappa A_{0}\text{,}
\label{bps2}
\end{equation}%
which must be solved together with the Gauss law (\ref{gg1}),%
\begin{equation}
\frac{d^{2}A_{0}}{dr^{2}}+\frac{1}{r}\frac{dA_{0}}{dr}+\kappa B=\frac{g^{2}h%
}{2}A_{0}\sin ^{2}\alpha \text{.}  \label{bps3}
\end{equation}

In addition, the reader can verify that the solutions for $n<0$
(anti-vortices) can be obtained directly from the ones for $n>0$ (vortices)
via the transformations $n\rightarrow -n$, $\alpha \rightarrow \alpha $, $%
A\rightarrow -A$ and $A_{0}\rightarrow -A_{0}$.

\begin{figure}[tbp]
\centering\includegraphics[width=8.5cm]{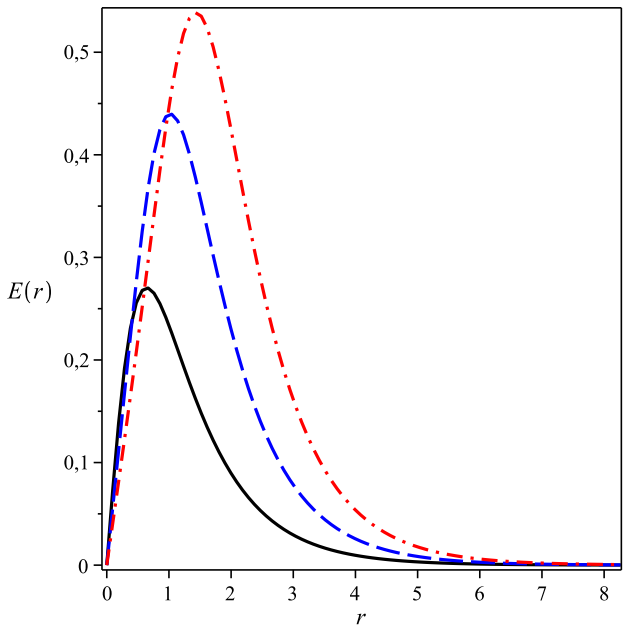}
\caption{Numerical solutions to the electric field $E(r)$. Conventions as in
the Fig. 1, $E(r=0)$ and $E(r\rightarrow \infty )$ vanishing.}
\end{figure}

Moreover, another quantity commonly referred when studying vortices is the
magnetic flux $\Phi _{B}$ they give rise. In the present case, the resulting
flux reads 
\begin{equation}
\Phi _{B}=2\pi \int_{0}^{\infty }\!\!rB\left( r\right) dr=\frac{4\pi }{g}n%
\text{.}  \label{ai7}
\end{equation}%
In this case, whether we consider 
\begin{equation}
\rho (r)=\frac{g^{2}h}{2}A_{0}\sin ^{2}\alpha
\end{equation}%
as the electric charge density, the magnetic flux $\Phi _{B}$ (\ref{ai7})
can be verified to be proportional to the total electric charge $\mathcal{Q}$%
, i.e. 
\begin{equation}
\mathcal{Q}=2\pi \int_{0}^{\infty }\!\!r\rho (r)dr=\kappa \Phi _{B}\text{,}
\label{cargax}
\end{equation}%
where we have integrated the Gauss law (\ref{gg1}).

\section{First-order solutions}

We now proceed the analysis of the differential equations (\ref{bps11}), (%
\ref{bps2}), (\ref{bps3}) themselves in view of the boundary conditions (\ref%
{ai00}) and (\ref{ai01}). It is useful to check \emph{a priori} whether the
solutions attain the boundary values (\ref{ai00}) and (\ref{ai01})
correctly. In order to verify such a convergence, we proceed to solve the
differential equations (\ref{bps11}), (\ref{bps2}) and (\ref{bps3}) around
those values. By considering $n>0$, we calculate the approximate solutions
near the origin to be 
\begin{eqnarray}
\alpha (r) &\approx &C_{n}r^{n}\text{,}  \label{a1a} \\[0.12in]
A(r) &\approx &\frac{g}{2}\left( gh-\kappa A_{0}(0)\right) r^{2}\text{,} 
\notag
\end{eqnarray}%
\begin{equation}
A_{0}(r)\approx A_{0}(0)-\frac{\kappa }{4}\left( gh-\kappa A_{0}(0)\right)
r^{2}\text{,}  \label{a2}
\end{equation}%
where $C_{n}$ and $A_{0}(0)\equiv A_{0}(r=0)$ are positive constants to be
determined numerically for each value of the vorticity $n$, the last
solution verifying the boundary value for $A_{0}(r=0)$ appearing in (\ref%
{ai00}).

A similar procedure reveals that the asymptotic profiles are (here, the
positive constant $C_{\infty }$ also depends on $n$ ) 
\begin{eqnarray}
\alpha (r) &\approx &\frac{\pi }{2}-C_{\infty }e^{-M_{\alpha }r}\text{,~}
\label{a3} \\[0.12in]
A(r) &\approx &2n-2MC_{\infty }re^{-M_{A}r}\text{,}  \notag
\end{eqnarray}%
\begin{equation}
A_{0}(r)\approx \frac{2M}{g}C_{\infty }e^{-Mr}\text{,}  \label{a4}
\end{equation}%
where the parameters 
\begin{equation}
M_{\alpha }=M_{A}=M=\frac{1}{2}\left( \sqrt{2g^{2}h+\kappa ^{2}}-\left\vert
\kappa \right\vert \right)
\end{equation}%
represent the masses of the corresponding bosons (the relation $M_{\alpha
}/M_{A}=1$ defining the BPS point). In this case, it is easy to verify that,
in the limit $g\gg \kappa $, the masses behave like those arising from a
gauged $CP(2)$ theory endowed by the Maxwell term only \cite{loginov}, 
\begin{equation}
M\approx g\sqrt{\frac{h}{2}}\text{.}
\end{equation}%
On the other hand, when $g\ll \kappa $, the masses mimic the value obtained
from a gauged $CP(2)$ model in the presence of the Chern-Simons action alone 
\cite{caduCS}:%
\begin{equation}
M\approx \frac{g^{2}h}{2\kappa }\text{.}
\end{equation}

In what follows, we proceed the numerical integration of the equations (\ref%
{bps11}), (\ref{bps2}), (\ref{bps3}) according the conditions (\ref{ai00})
and (\ref{ai01}). We solve the corresponding system by means of
finite-difference algorithm, from which we depict the resulting profiles in
the figures 1, 2, 3, 4, 5 and 6. Here, we choose $h=\kappa =1$ and $g=2$,
plotting the solutions for $n=1$ (solid black line), $n=2$ (dashed blue
line) and $n=3$ (dotted red line).

\begin{figure}[tbp]
\centering\includegraphics[width=8.5cm]{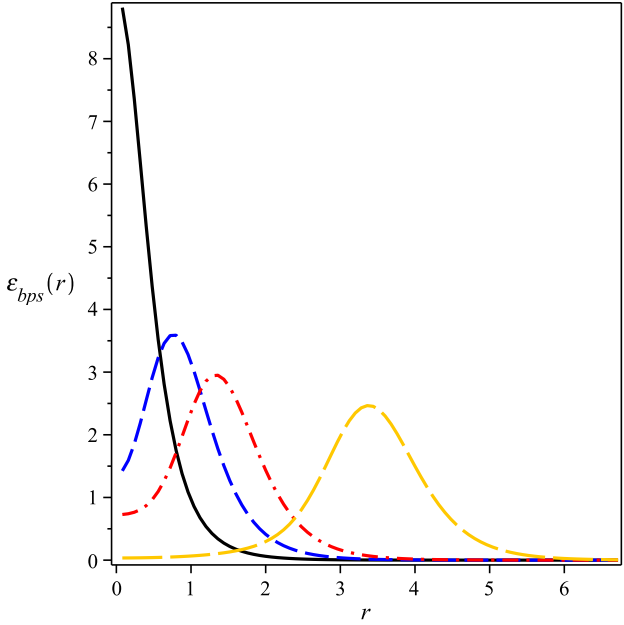}
\caption{Numerical solutions to the energy distribution $\protect\varepsilon %
_{bps}(r)$. Conventions as in the Fig. 5, the resulting structures being
similar to the magnetic ones.}
\end{figure}

The figures 1 and 2 show the numerical solutions to the profile functions $%
\alpha (r)$ and $A(r)$,respectively, from which we see that these two
solutions reach the boundary values in a monotonic way, as expected. In
particular, the gauge function obeys $A(r\rightarrow \infty )\rightarrow 2n$.

In the Figure 3, we depict the numerical profile of the scalar potential $%
A_{0}(r)$, the corresponding structures being lumps centered at $r=0$. It is
interesting to note how $A_{0}(0)\equiv A_{0}(r=0)$ depends on $n$, its
value increasing as the vorticity itself increases until reaching the
maximum value%
\begin{equation}
A_{0}^{max}(r=0)=\frac{gh}{\kappa }\text{,}
\end{equation}%
for sufficiently large values of $n$ (see the Figure 4).

The Figure 5 presents the solutions for the magnetic field $B(r)$, including
the ones obtained for $n=7$ (long-dashed orange line) and $n=12$ (dotted
green line). In this case, for $n=1$ , the profile is a well-defined lump
centered at the origin. On the other hand, for $n>1$, the solutions stand
for rings also centered at $r=0$, their amplitudes (radii) decreasing
(increasing) as the vorticity increases.

\begin{figure}[tbp]
\centering\includegraphics[width=8.5cm]{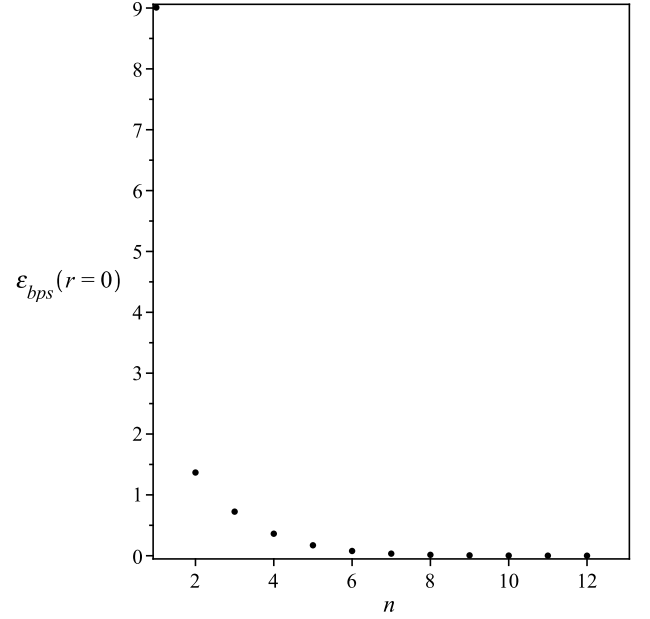}
\caption{The dependence of $\protect\varepsilon _{bps}(0)$ with the
vorticity: for $n=1$, one gets $\protect\varepsilon _{bps}(r=0)=2( C_{1})
^{2}h+b_0^2$.}
\end{figure}

It is particularly interesting to consider the solution for$B(r)$ coming
from (\ref{a1a}). It reads 
\begin{equation}
B(r)\approx b_{0}+\frac{1}{4}\left( \kappa ^{2}b_{0}-2\left( C_{1}\right)
^{2}gh\right) r^{2}\text{,}
\end{equation}%
for $n=1$, and 
\begin{equation}
B(r)\approx b_{0}+\frac{\kappa ^{2}b_{0}}{4}r^{2}\text{,}
\end{equation}%
for $n\geq 2$, where we have defined 
\begin{equation}
b_{0}=gh-\kappa A_{0}(0)\text{.}
\end{equation}

In this sense, we get that value of $B(r=0)$, i.e.%
\begin{equation}
B(r=0)=b_{0}=gh-\kappa A_{0}(0)\text{,}
\end{equation}%
which explains the behavior of the magnetic field at the origin: as $n$\
increases, the value of $A_{0}(r=0)$\ increases too (see the Figure 4),
whilst the difference $b_{0}=gh-\kappa A_{0}(0)$ decreases, see the Figure
6. In particular, for sufficiently large values of the vorticity, one gets
that $A_{0}(r=0)\rightarrow gh/\kappa $\ and $B(r=0)=b_{0}\rightarrow 0$.

The solutions to the electric field $E(r)$\ appear in the Figure 7, from
which we see that the resulting configurations are rings centered at $r=0$,
their amplitude and radii increasing as the vorticity increases, with both $%
E(r=0)$\ and $E(r\rightarrow \infty )$\ vanishing.

We depict the numerical profiles to the energy distribution $\varepsilon
_{bps}(r)$\ in the Figure 8, from which see that the resulting structures
are similar to the magnetic ones, i.e. for $n=1$ ($n>1$) the solution is a
lump (ring) centered at $r=0$. Thus, it is instructive to study the behavior
of $\varepsilon _{bps}(r)$ near the origin which results to be%
\begin{equation}
\varepsilon _{bps}(r)\approx 2\left( C_{1}\right) ^{2}h+\left( b_{0}\right)
^{2}+\varepsilon _{2}r^{2}\text{,}
\end{equation}%
from $n=1$, and%
\begin{equation}
\varepsilon _{bps}(r)\approx \left( b_{0}\right) ^{2}+\frac{3}{4}\left(
b_{0}\kappa \right) ^{2}r^{2}\text{,}
\end{equation}%
for $n\geq 2$. Here, we have defined%
\begin{equation}
\varepsilon _{2}=\frac{3}{4}\left( b_{0}\kappa \right) ^{2}-\frac{2}{3}%
\left( C_{1}\right) ^{4}h+\frac{gh}{2}\left( C_{1}\right) ^{2}\left[ \left(
A_{0}(0)\right) ^{2}g-4b_{0}\right] \text{,}
\end{equation}%
for the sake of convenience.

We therefore note that, as $n$\ increases, $\varepsilon _{bps}(r=0)$\
vanishes. In particular, for $n=1$, one gets that $\varepsilon
_{bps}(r=0)=2\left( C_{1}\right) ^{2}h+\left( b_{0}\right) ^{2}$, see the
Figure 9.

\section{Final comments and perspectives}

We have shown the existence of radially symmetric first-order solitons in a
gauged $CP(2)$ model endowed by the Maxwell and the Chern-Simons terms
simultaneously.

The point to be raised here is that, in order to implement the Bogomol'nyi
prescription correctly, we have introduced a neutral scalar field into the
original Maxwell-Chern-Simons model (\ref{ai1}), such a modification being a
well-known procedure usually performed when the dynamics of the gauge sector
is controlled by the composite Maxwell-Chern-Simons action. Such a procedure
also works in the Lorentz-violating extensions of the simplest
Maxwell-Chern-Simons-Higgs theory and of the $O(3)$-sigma model with the
Maxwell-Chern-Simons term \cite{casana1,Guillermo,Claudio, Claudio1}.

We also have calculated the self-interacting potential engendering
self-duality, from which we also have determined the lower-bound for the
overall energy (i.e. the Bogomol'nyi bound) and the corresponding
first-order differential equations supporting topological solitons.
Moreover, we have verified that the resulting Bogomol'nyi bound is
proportional to the topological charge of the model, both ones being
quantized according the vorticity $n$, as expected. In addition, we have
rewritten the total electric charge $\mathcal{Q}$ in terms of the magnetic
flux $\Phi _{B}$, which is also quantized, see the equations (\ref{ai7}) and
(\ref{cargax}).

We are now studying the existence of nontopological solitons in a gauged $%
CP(2)$ scenario in the presence of the Chern-Simons and of the
Maxwell-Chern-Simons terms, from which we hope interesting results to be
reported in a future contribution.

\begin{acknowledgments}
This work was supported by the Brazilian Government via the Conselho
Nacional de Desenvolvimento Cient{\'\i}fico e Tecnol\'ogico (CNPq), the
Coordena\c{c}\~ao de Aperfei\c{c}oamento de Pessoal de N{\'\i}vel Superior
(CAPES), and the Funda\c{c}\~ao de Amparo \`a Pesquisa e ao Desenvolvimento
Cient{\'\i}fico e Tecnol\'ogico do Maranh\~ao (FAPEMA). In particular, RC
thanks the support from the grants CNPq/306385/2015-5,
FAPEMA/Universal-00782/15 and FAPEMA/Universal-01131/17, NHGG acknowledges
the full support from CAPES (postgraduate scholarship), and EH thanks the
support from the grants CNPq/307545/2016-4 and CNPq/449855/2014-7.
\end{acknowledgments}

\end{document}